# Edelstein effect induced superconducting diode effect in inversion symmetry breaking MoTe$_2$ Josephson junctions

P. B. Chen[1,2], G.Q. Wang[2], B. C. Ye[2,3], J. H. Wang[2], L. Zhou[2], Z. Z. Tang[2], L. Wang[4], J. N. Wang[3], W. Q. Zhang[5,6], J. W. Mei[2,6], W. Q. Chen[2,6] and H. T. He[2,6*]

[1]*Department of physics, Harbin Institute of Technology, Harbin, 150001, China*
[2]*Department of Physics, Southern University of Science and Technology, Shenzhen 518055, China*
[3]*Department of Physics, The Hong Kong University of Science and Technology, Clear Water Bay, Hong Kong, China*
[4]*Shenzhen Institute for Quantum Science and Engineering, Southern University of Science and Technology, Shenzhen, 518055, China*
[5]*Department of Materials Science and Engineering, Southern University of Science and Technology, Shenzhen 518055, China*
[6]*Shenzhen Key Laboratory for Advanced Quantum Functional Materials and Devices, Southern University of Science and Technology, Shenzhen 518055, China*
[*]*Corresponding author email address: heht@sustech.edu.cn*

**Abstract.**
Superconducting diode effect (SDE) with nonreciprocal supercurrent transport has attracted intense attention recently, not only for its intriguing physics, but also for its great application potential in superconducting circuits. It's revealed in this work that planar Josephson junctions (JJs) based on type-II Weyl semimetal (WSM) MoTe$_2$ can exhibit a prominent SDE due to the emergence of asymmetric Josephson effect (AJE) in perpendicular magnetic fields. The AJE manifests itself in a very large asymmetry in the critical supercurrents with respect to the current direction. The sign of this asymmetry can also be effectively modulated by the external magnetic field. Considering the special noncentrosymmetric crystal symmetry of MoTe$_2$, this AJE is understood in terms of the Edelstein effect, which induces a nontrivial phase shift in the current phase relation of the junctions. Besides these, we further demonstrate the rectification of supercurrent in such MoTe$_2$ JJs with the rectification efficiency up to 50.4%, unveiling the great application potential of WSMs in superconducting electronics.

Superconducting diode effect (SDE), which was first observed in noncentrosymmetric [Nb/V/Ta]$_n$ superlattices [1], refers to a nonreciprocal transport behavior where the supercurrent only flows in one direction, but vanishes in the opposite current direction. It's not only an intriguing phenomenon in physics, but also of great importance to the development of low-dissipation superconducting electronics. Therefore, intense interest has been triggered recently in searching for new SDE systems [1-8], as well as exploring the physical mechanism of it [9-19]. Latest studies have revealed that noncentrosymmetric Josephson junctions (JJs) can be a promising platform to realize this SDE due to the emergence of asymmetric Josephson effect (AJE) [3-8]. In such JJs, the critical Josephson supercurrents ($I_c$) are asymmetric with respect to the current direction, *i.e*, $I_c^+ \neq |I_c^-|$, where superscripts $+$ & $-$ represent the positive and negative current directions, respectively. The SDE would be expected to occur with $-I_c^- < I < I_c^+$ if $I_c^+ > -I_c^-$ or $I_c^- < I < -I_c^+$ if $I_c^+ < -I_c^-$. Such AJE-induced SDEs have been reported in the NbSe$_2$/Nb$_3$Br$_8$/NbSe$_2$ tunneling vertical JJ [3], the highly transparent Al/InAs/Al and Nb/Pt/Nb planar JJs [4,5], and the Nb/NiTe$_2$/Nb and Nb/InSb/Nb underdamped planar JJ [6,7].

ARPES studies have shown that MoTe$_2$ is a type-II Weyl semimetal (WSM) with strongly tilted bulk Weyl cones and surface Fermi arcs connecting the Weyl nodes [20]. It has many exotic physical properties, such as the extremely large magnetoresistance [21-23], the possible two-gap superconductivity [24,25], and the edge supercurrent [26]. Especially, noncentrosymmetric MoTe$_2$ is expected to exhibit the Edelstein effect [27,28]. Due to the close connection between the Edelstein effect and the AJE [29,30], MoTe$_2$-based Josephson junctions are thus believed to exhibit the AJE and have a great application potential in the superconducting diodes. But up to now, no such Edelstein effect induced AJE in MoTe$_2$ have been reported yet.

In this work, we have successfully fabricated the Nb/MoTe$_2$/Nb planar JJs and systematically studied the critical supercurrents $I_c$ as a function of the perpendicular magnetic field ($B$). Due to the time reversal symmetry, the magnitude of the supercurrent is unchanged with both the directions of the current and field reversed. But in sharp contrast with the standard Fraunhofer pattern, the inversion symmetry breaking in MoTe$_2$ leads to asymmetric critical supercurrents, *i.e.*, $|I_c^\pm(B)| \neq |I_c^\mp(B)|$ & $I_c^\pm(B) \neq I_c^\pm(-B)$. The emergence of an anomalous phase in the current phase relation of the Josephson junction due to the Edelstein effect is believed to be the main physical origin of this critical current asymmetry. It's also found that the sign of the critical supercurrent difference between opposite directions, *i.e.*, $\Delta I_c(B) = I_c^+(B) - |I_c^-(B)|$, can be effectively modulated by the external magnetic field. Based on this field-tunable AJE, we further demonstrate the SDE in these MoTe$_2$-based JJs with the largest rectification efficiency obtained so far, paving the way for future application of WSMs in low-dissipation superconducting electronics.

The MoTe$_2$ flakes were exfoliated from the bulk crystals (Section S1, Supporting Information) and then transferred to the Si substrate with a top layer of 280 nm insulating SiO$_2$. Using the e-beam lithography and pulsed laser deposition (PLD) techniques, a pair of superconducting Nb electrodes were fabricated on the flakes, forming the Nb/MoTe$_2$/Nb planar Josephson junctions. In order to enhance the electrical transparency of the Nb/MoTe$_2$ contact, an Ar ion source has been implemented to clean the surface of MoTe$_2$ before in situ depositions of the Nb electrodes by PLD. Fig. 1 (a) shows the image of a typical Nb/MoTe$_2$/Nb JJ, with the gap ($L$) between the two Nb electrodes of about 300 nm. The transport properties of the junctions were investigated in a dilution refrigerator under different temperature and field conditions. The differential resistance curves were measured using Keithley 6221 AC/DC current source and 2182A nanovoltmeter.

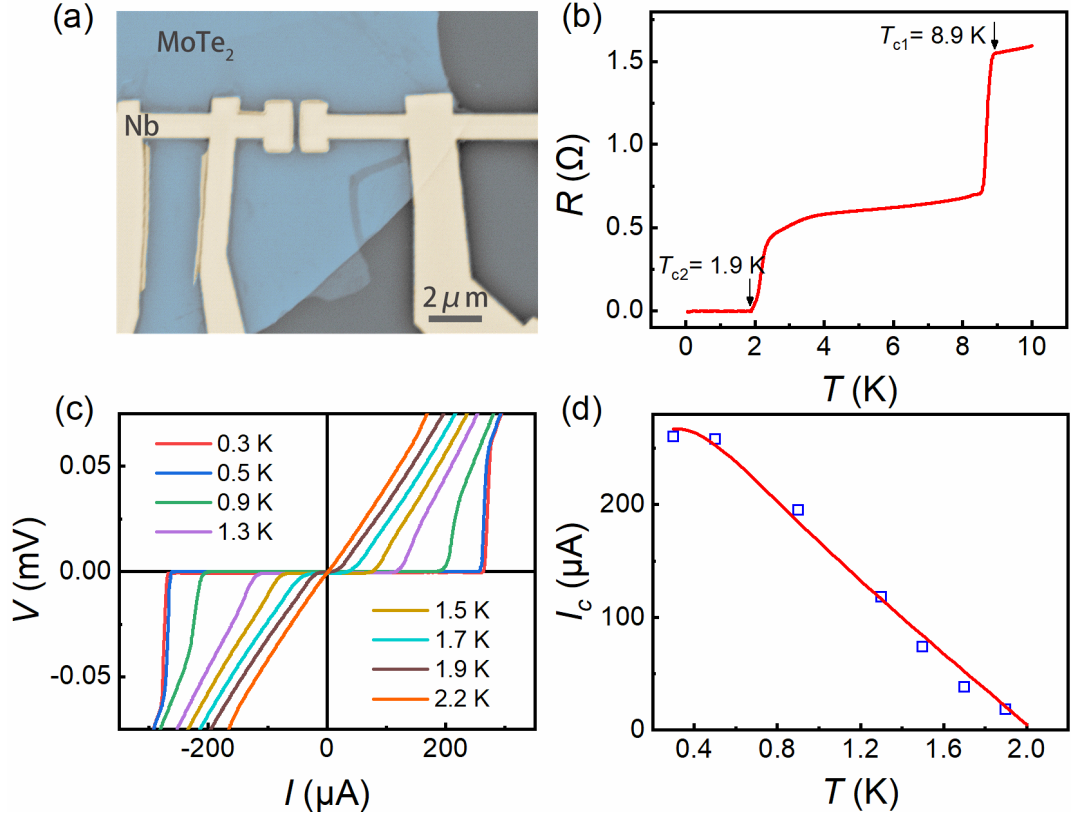

**Figure 1**. (a) An image of the planar MoTe$_2$ Josephson junction. The thickness ($t$) of the MoTe$_2$ flake is 33 nm and the junction has a width ($W$) of 2 μm and a length ($L$) of 300 nm. (b) The temperature dependence of the resistance of the Josephson junction. The Nb superconducting transition occurs around $T_{c1}$=8.9 K and the Josephson supercurrents flow below $T_{c2}$=1.9 K. (c) The voltage ($V$)-current ($I$) curves at different temperatures ranging from 0.3 K to 2.2 K in zero magnetic field. (d) The temperature dependence of the critical supercurrent $I_c$. The data can be well fitted by the Zaikin-Zharkov theory, as indicated by the red fitting curve.

Fig. 1 (b) shows the resistance ($R$) of the junction as a function of the temperature ($T$). With decreasing temperatures, two sharp resistance drops are successively observed. The first resistance drops around $T_{c1}$=8.9 K arises from the superconducting transition of the Nb electrodes, while the second one indicates the flowing of supercurrent in the junction, since the resistance drops to zero below $T_{c2}$=1.9 K. Fig. 1 (c) shows the voltage ($V$)-current ($I$) characteristics of the junction obtained at different temperatures. Note that in measuring the $V(I)$ curves in Fig. 1 (c), the current was always swept from zero to positive or negative values in order to suppress any possible heating effect (Fig. S2, Supporting Information). Compared with the $V(I)$ curves above $T_{c2}$, the curves below $T_{c2}$ clearly exhibit a zero-resistance state when the bias current lies between $I_c^+$ and $I_c^-$. $I_c^+$ & $I_c^-$ represent the critical Josephson supercurrents in the positive and negative current directions, respectively. Note that $I_c^+ = |I_c^-|$ in zero magnetic fields. From the $V(I)$ curves in Fig. 1 (c), we can obtain the temperature dependence of the critical supercurrent $I_c$ as shown in Fig. 1 (d).

We have performed the longitudinal and Hall resistivity measurement of the MoTe₂ flakes, from which the carrier density ($n$) and mobility ($\mu$) are derived to be $4.849 \times 10^{20}$ cm⁻³ and 1115 cm²V⁻¹s⁻¹ at 2 K. Due to the coexistence of multiple electron and hole pockets in MoTe₂ [20,31,32], it's difficult for us to directly derive the mean free path ($l$) of MoTe₂. But in a previous transport study of Shubnikov-de Hass oscillations in MoTe₂ flakes with $n=2.3 \times 10^{20}$ cm⁻³ and $\mu=3900$ cm²V⁻¹s⁻¹, a mean free path of 105 nm was obtained [33]. Considering the lower mobility of our MoTe₂ flakes, the mean free path is believed to be smaller than the distance between Nb electrodes of our device ($L \sim 300$ nm). Therefore, we use the Zaikin-Zharkov theory developed for diffusive Josephson junctions with $L > l$ to fit the obtained experimental data $I_c(T)$ [34,35]. It's seen that the data can be well described by the Zaikin-Zharkov theory, as indicated by the red fitting curve with only two fitting parameters $\Delta^*$ & $D$ in Fig. 1 (d). The fitting yields the superconducting gap $\Delta^* \sim 0.145$ meV and the diffusion coefficient $D \sim 0.155$ m²/s. It's worth pointing out that as the Josephson supercurrent only appears below $T_{c2}$, it is more appropriate to ascribe $\Delta^*$ to the proximity induced superconducting gap in MoTe₂ just below the Nb electrodes, rather than the gap of the Nb electrode ($1.76\,k_B T_{c1} \sim 1.3$ meV). Similar behaviors have been discussed in previous studies of planar Josephson junctions [4,36,37].

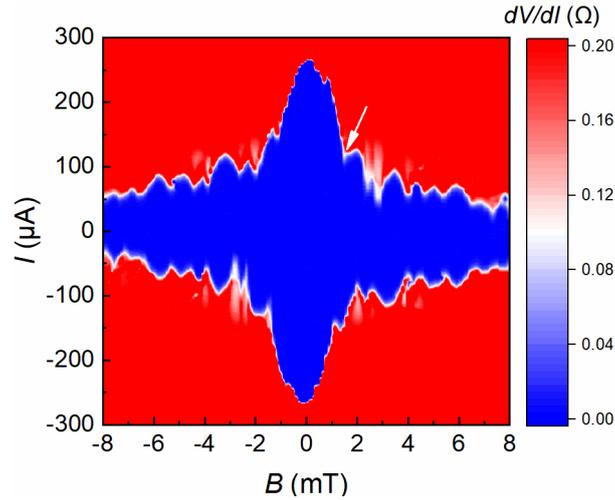

**Figure 2.** 2D color map of the differential resistance versus the current bias and magnetic field at 0.3 K, exhibiting oscillatory critical supercurrents due to the interference of Josephson supercurrent in magnetic fields.

The presence of AJE in Nb/MoTe₂/Nb junctions is revealed by investigating the supercurrents in perpendicular magnetic fields ($B$). Fig. 2 shows the field and current bias dependence of the junction differential resistance $\frac{dV}{dI}$ at $T$=0.3 K. Similar to Fig. 1 (c), the current bias was always swept from zero to positive or negative values. In the blue region, $\frac{dV}{dI} \sim 0$, *i.e.*, the supercurrent flows through the junction. It's apparent that the critical supercurrent $I_c$, the boundary of the blue region, exhibits a characteristic oscillatory behavior as the field changes, reflecting the interference of supercurrents in perpendicular fields. But different from the standard Fraunhofer patterns [38], the oscillations are not periodic, hindering the accurate determination of the effective junction area $A$ from the oscillation period $\Delta B$ via $A = W(L + 2\lambda) = \frac{\Phi_o}{\Delta B}$, where $\Phi_o$ is the flux quantum, and $W$, $L$, and $\lambda$ represents the junction width, length, and the magnetic field penetration depth of Nb, respectively. As a rough estimation, we take the field difference between the first minima in $I_c(B)$ as indicated by the arrow in Fig. 2 and the main peak as $\Delta B$ and obtain the effective junction area $A$~1.477 µm². As $W$=2 µm & $L$=300 nm for the junction shown in Fig. 1 (a), the value of $\lambda$ can be calculated to be about 219 nm, consistent with previous studies [39].

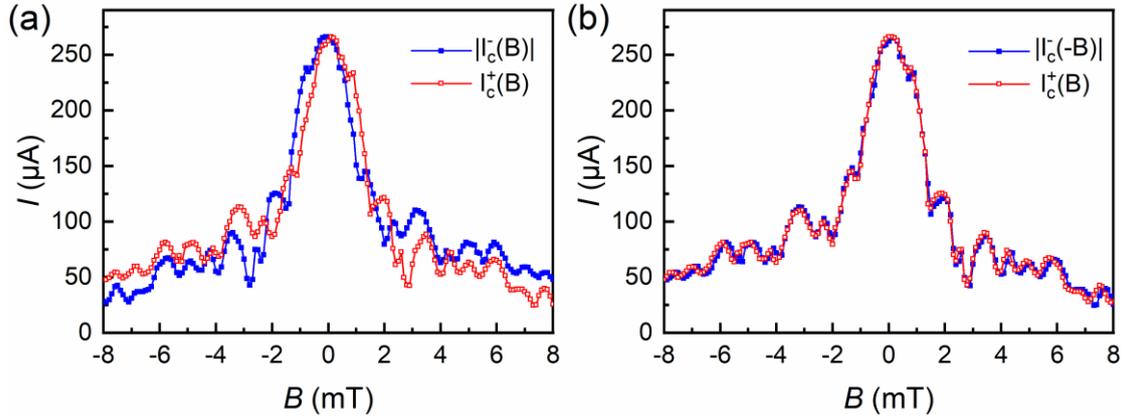

**Figure 3.** (a) Field dependence of the magnitudes of critical supercurrents in positive ($I_c^+$) and negative ($|I_c^-|$) current directions at 0.3 K. (b) By reversing the field direction of the negative critical supercurrents, the resultant $|I_c^-(-B)|$ curve collapses nicely on the $I_c^+(B)$ curve, indicating the symmetry $|I_c^\pm(B)| = |I_c^\mp(-B)|$.

More interestingly, the interference pattern is asymmetric with respect to the field or the current direction, *i.e.*, $I_c^\pm(B) \neq I_c^\pm(-B)$ or $|I_c^\pm(B)| \neq |I_c^\mp(B)|$. For clarity, the $I_c^+(B)$ and $|I_c^-(B)|$ curves are plotted in Fig. 3 (a), which clearly demonstrates the above asymmetry. Such an AJE is very different from the standard symmetric Fraunhofer pattern usually observed in SIS junctions [38]. But it's noted that the critical supercurrent is symmetric if both the current and field directions are reversed simultaneously. To illustrate this, we plot the $I_c^+(B)$ and $|I_c^-(-B)|$ curves in Fig. 3 (b). These two curves overlap with each other nicely, revealing that the interference pattern preserves the symmetry $|I_c^\pm(B)| = |I_c^\mp(-B)|$, which is a direct manifestation of the time reversal symmetry [10]. Fig. 3 (b) also indicates that the oscillotory behavior of the $I_c(B)$ curve is a real physical phenomenon, but not an artifact. Note that similar results are also observed in other MoTe₂ planar Josephson junctions (Fig. S3, Supporting Information).

To further study the AJE quantitatively, the difference $\Delta I_c(B)$ between the $I_c^+(B)$ and $|I_c^-(B)|$ curves has been calculated at different temperatures and plotted in Fig. 4. Since $|I_c^\pm(B)| = |I_c^\mp(-B)|$, all the $\Delta I_c(B)$ curves in Fig. 4 exhibits an antisymmetry, i.e., $\Delta I_c(-B) = -\Delta I_c(B)$. The maximum $|\Delta I_c|$, i.e., the most pronounced AJE, appears around $\pm 1$ mT and can be as large as 62 µA, which is among the largest values of $|\Delta I_c|$ obtained so far in JJ-based superconducting diodes [3-8]. With increasing fields, the magnitude of $\Delta I_c$ tends to decrease due to the suppression of the supercurrent by fields. It's also interesting to see that the sign of $\Delta I_c$ can be modulated by the magnetic field. For example, the $\Delta I_c(B)$ curve obtained at 0.3 K clearly alternates between positive and negative signs with increasing fields, as shown in Fig. 4. This result clearly demonstrates the tunability of AJE with the magnetic field. As the temperature increases, the AJE weakens gradually as expected, especially with $|B| > 1.5$ mT.

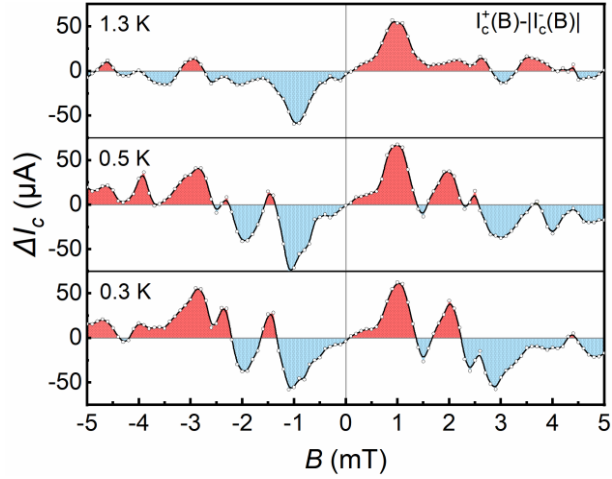

**Figure 4.** Field dependence of the asymmetry in critical supercurrents defined by $\Delta I_c = I_c^+(B) - |I_c^-(B)|$ at three different temperatures. The field regions with positive and negative signs of $\Delta I_c$ are shaded with the light red and blue colors, respectively.

AJE has attracted intense attention recently due to its great potential for applications in superconducting diodes. It has been observed in vertical or planar Josephson junctions made from the InAs quantum wells, the InSb nanoflakes, the heavy metal Pt, and the type-II Dirac semimetal NiTe$_2$ [4-7]. In previous studies of the planar Josephson junctions, an extra in-plane magnetic field is found crucial to the occurrence of AJE in addition to a perpendicular field [4-7,40]. It's this in-plane field that gives rise to an anomalous phase $\varphi_0$ in the current phase relation (CPR) of these planar Josephson junctions. Together with higher order harmonics, this phase-shifted non-sinusoidal CPR leads to the observed AJE [3-7]. But in our work, a perpendicular field alone is sufficient to the emergence of asymmetric critical currents as shown in Fig. 4, without any need to apply the in-plane field. In order to understand this discrepancy, we need to revisit the close connection between the $\varphi_0$-junction behavior and the Edelstein effect. The Edelstein effect refers to the phenomenon that the supercurrent $\vec{J}_s$ of a noncentrosymmetric superconductor (SC) with spin-orbit coupling (SOC) can induce a net spin magnetization $\vec{M}$. It's first proposed by V. M. Edelstein in polar SCs with Rashba SOC and then generalized to other noncentrosymmetric SCs belonging to the 18 gyrotropic groups [27,28]. Similar behavior can also occur in Josephson junctions, *i.e.*, the Josephson supercurrent through the weak-link with SOC will generate a spin magnetization via the Edelstein effect. If the spin magnetization has a component parallel to the external magnetic field $\vec{B}$ applied to the Josephson junction, *i.e.*, $\vec{M} \cdot \vec{B} \neq 0$, the coupling of it with the field will contribute a Zeeman-like energy term to the free energy of the junction. It will eventually give rise to a nontrivial phase $\varphi_0$ in the CPR of the junction [41]. As quasi-2D MoTe$_2$ belongs to the $C_{1v}$ point group, an in-plane supercurrent can induce an out-of-plane spin magnetization [28], as discussed in Section S2, Supporting Information. According to the above discussion, MoTe$_2$ Josephson junction in presence of a perpendicular magnetic field fulfills the condition $\vec{M} \cdot \vec{B} \neq 0$ and thus behaves like a $\varphi_0$-junction. It's this anomalous phase $\varphi_0$ induced in the CPR via the Edelstein effect, together with higher harmonics, that results in the asymmetric Josephson effect in our MoTe$_2$ Josephson junctions. In fact, just after the proposal of the Edelstein effect, it was shown by Edelstein that asymmetric critical currents can be realized in polar SCs via the Edelstein effect [29]. A latest theoretical study also points out the essential role of the Edelstein effect in understanding the Josephson diode effect [30]. To further justify the Edelstein effect induced AJE picture, we also fabricated Josephson junctions based on PdTe$_2$, an inversion symmetric type-II Dirac semimetal [42]. As it is centrosymmetric and does not belong to any one of the 18 gyrotropic point groups [28], no Edelstein effect is expected. Therefore, there will be no AJE in PdTe$_2$-based Josephson junctions, which is good agreement with the experimental results (Fig. S5, Supporting Information). All these thus strongly indicate that the microscopic mechanism for the AJE in MoTe$_2$ Josephson junctions is the Edelstein effect due to the inversion symmetry breaking in MoTe$_2$ [28].

Within this Edelstein effect induced AJE picture, we can now understand the difference between our work and previous studies of Josephson supercurrent diodes with Rashba-type SOC [4-7]. For such junctions, an in-plane supercurrent only induces an in-plane spin magnetization [28]. Therefore, an in-plane field is required to realize the $\varphi_0$-junction. Besides this, it's also noted that in these junctions the value of $\varphi_0$ is a constant once the in-plane field is fixed and won't change with the perpendicular field. But in our MoTe$_2$ Josephson junctions, the induced phase $\varphi_0$ is expected to change with the perpendicular field due to the coupling of the out-of-plane spin magnetization to the perpendicular field, which might explain the non-periodic oscillations in critical supercurrents shown in Fig. 2, as well as the field-tunable AJE in Fig. 4 (Section S3, Supporting Information). Although the anomalous phase shift $\varphi_0$ has been derived explicitly for JJs with Rashba-type SOC [40,41], no such calculation of $\varphi_0$ has been performed for JJs with the weak link belonging to the $C_{1v}$ point group. Therefore, our work calls for more theoretical efforts to quantitatively understand the observed phenomena in MoTe$_2$-based JJs.

We can also rule out the magnetochiral anisotropy as the main origin of the observed nonreciprocal transport phenomena. The magnetochiral anisotropy in noncentrosymmetric WSM has been revealed to have the form $\rho \propto \vec{I} \cdot \vec{B}$ [43]. With the current $\vec{I}$ perpendicular to the magnetic field $\vec{B}$ in our experiment, the magnetochiral anisotropy is expected to play a minor role in the observed AJE. We are aware that a recent theoretical study has predicted the occurrence of an intrinsic AJE in perpendicular fields in inversion symmetry breaking topological materials, such as type-II WSMs [10]. Such intrinsic AJEs were later reported in the study of edge superconductivity in few-layer type-II WSM WTe$_2$ [44,45]. But they were found to vanish in edge-untouched devices [44]. As the Nb electrodes of our junctions don't touch the edges of the MoTe$_2$ flake as shown in Fig. 1 (a), this mechanism is also unlikely responsible for the observed AJE in this work. Besides, only standard Fraunhofer pattern is observed in PdTe$_2$ Josephson junctions fabricated with the same nanofabrication process flow as the MoTe$_2$ junctions (Fig. S5, Supporting Information). This observation further excludes the possibility of non-uniform supercurrent distribution in our edge-untouched MoTe$_2$ junctions.

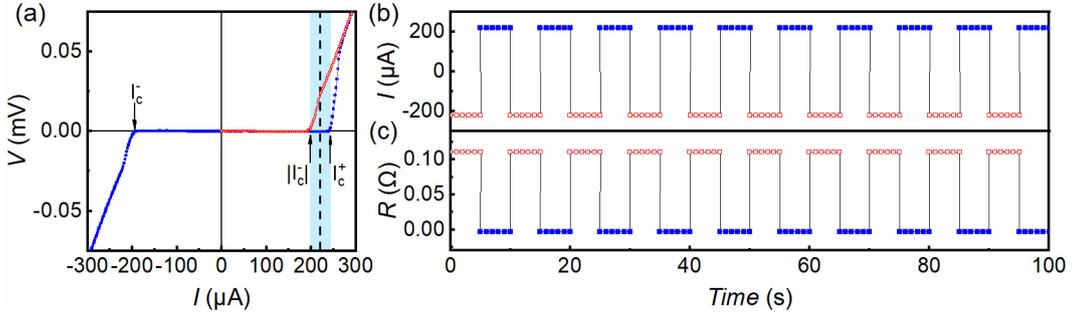

**Figure 5.** (a) The $V(I)$ curve measured in perpendicular magnetic field of 0.8 mT at 0.3 K, with critical supercurrents $I_c^+$ & $I_c^-$ as indicated by arrows. For comparison, the absolute values of the critical supercurrents obtained in the negative current direction are also plotted. In the shaded region with $|I_c^-| < I < I_c^+$, the SDE occurs with the supercurrent only flowing in the positive direction. The dashed line indicates the current applied in the rectification experiment. (b) Periodically switching of the bias current between 220 μA and -220 μA and (c) the corresponding change in the junction resistance, which clearly alternates between supercurrent and resistive states, as indicated by the blue solid and red open squares, respectively.

The observation of AJE indicates that there exists a certain current region where the Josephson supercurrent only flows in one direction. To see this behavior more clearly, the corresponding $V(I)$ curve of the junction at 0.8 mT is shown by the blue curve in Fig. 5 (a), with the critical supercurrents $I_c^+$ & $I_c^-$ indicated by arrows. For comparison, the absolute values of the data obtained in the negative bias direction are also plotted as the red curve in Fig. 5 (a). It's thus obvious that the supercurrent only flows in the positive direction when $|I_c^-| < I < I_c^+$. Once the current direction is reversed, no supercurrent would flow, *i.e.*, the junction will be in a resistive state. Such a non-reciprocal phenomenon gives rise to the so-called superconducting diode effect [1-19]. In order to demonstrate this SDE, we have further performed the following switching experiment. We first fix the magnitude of current at 220 μA, as indicated by the vertical dashed line in Fig. 5 (a). We then switch the current direction periodically as shown in Fig. 5 (b). Due to the AJE, it's seen in Fig. 5 (b) that the junction alternates between the supercurrent and normal resistive states periodically.

These two states are indicated by blue solid and red open squares in Fig. 5 (b), respectively. This switching experiment clearly demonstrates the realization of SDE in our MoTe$_2$ Josephson junctions with broken inversion symmetry. We also calculate the rectification efficiency of the junctions, which is defined by $\frac{I_c^+ - |I_c^-|}{I_c^+ + |I_c^-|}$. The maximum efficiency can reach up to 50.4 % (Fig. S6, Supporting Information). To the best of our knowledge, this is the largest value obtained so far in superconducting diodes [1-8]. Here we cannot compare the rectification efficiency between our MoTe$_2$ and previous WTe$_2$-based Josephson junctions, since no such analysis or SDE experiment has been performed in the study of WTe$_2$ Josephson junctions [43,44].

In conclusion, we have revealed the emergence of a prominent asymmetric Josephson effect in MoTe$_2$-based Josephson junctions, which is highly field-tunable and ascribed to the Edelstein effect in MoTe$_2$. Due to this AJE, it's demonstrated that the junction can be operated as an efficient superconducting diode to rectify the Josephson supercurrent. Compared with previous superconducting diodes, it exhibits the largest rectification efficiency and high magnetic-field-tunability, without the need to apply an extra in-plane magnetic field due to the special crystal symmetry of MoTe$_2$. Our work not only verifies experimentally a new microscopic mechanism for SDE, *i.e.*, the Edelstein effect induced AJE, but also shows the great application potential of WSM in superconducting electronics.

**Acknowledgement**

This work was supported by the National Key Research and Development Program of China (No. 2022YFA1403700), the Natural Science Foundation of Guangdong Province (No. 2021A1515010046), the Science, Technology and Innovation Commission of Shenzhen Municipality (No. GXWD20201230110313001 & No. ZDSYS20190902092905285), Guangdong province 2020KCXTD001, and the Center for Analysis and Test of Southern University of Science and Technology.

# Supporting Information

# Edelstein effect induced superconducting diode effect in inversion symmetry breaking MoTe$_2$ Josephson junctions

P. B. Chen[1,2], G.Q. Wang[2], B. C. Ye[2,3], J. H. Wang[2], L. Zhou[2], Z. Z. Tang[2], L. Wang[4], J. N. Wang[3], W. Q. Zhang[5,6], J. W. Mei[2,6], W. Q. Chen[2,6] and H. T. He[2,6*]

[1]*Department of physics, Harbin Institute of Technology, Harbin, 150001, China*
[2]*Department of Physics, Southern University of Science and Technology, Shenzhen 518055, China*
[3]*Department of Physics, The Hong Kong University of Science and Technology, Clear Water Bay, Hong Kong, China*
[4]*Shenzhen Institute for Quantum Science and Engineering, Southern University of Science and Technology, Shenzhen, 518055, China*
[5]*Department of Materials Science and Engineering, Southern University of Science and Technology, Shenzhen 518055, China*
[6]*Shenzhen Key Laboratory for Advanced Quantum Functional Materials and Devices, Southern University of Science and Technology, Shenzhen 518055, China*

[*]Corresponding author email address: heht@sustech.edu.cn

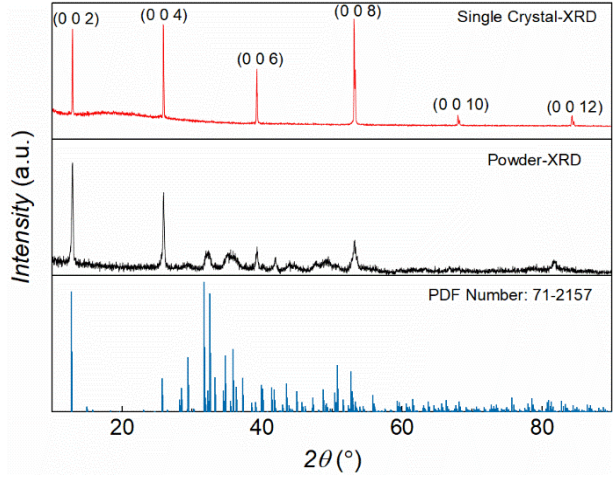

**Figure S1.** Single crystal and Powder X-ray diffraction spectra of $1T'$-MoTe$_2$ at room temperature.

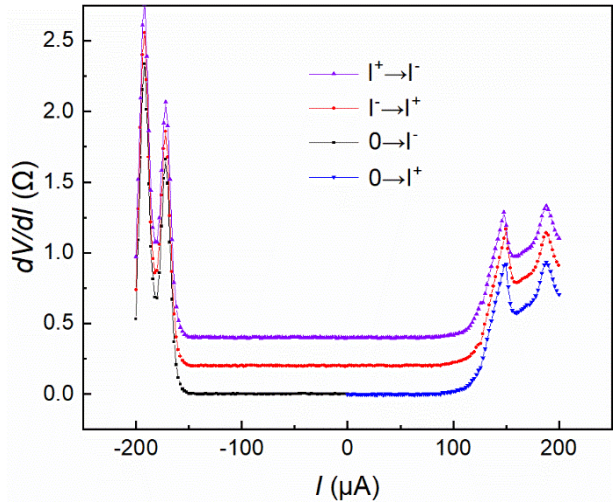

**Figure S2.** The differential resistance curves of the Josephson junction at $T$=0.5 K and $B$=0.5 mT. The curves are vertically shifted for clarity. The current sweeping is done from positive to negative currents ($I^+ \to I^-$) for the purple curve, from negative to positive currents ($I^- \to I^+$) for the red curve, from zero to negative currents ($0 \to I^-$) for the black curve, and from zero to positive currents ($0 \to I^+$) for the blue curve, respectively. It's clear that the obtained differential resistance curves are insensitive to the sweeping direction of currents, indicating the overdamped nature of the Josephson junction and the negligible heating effect.

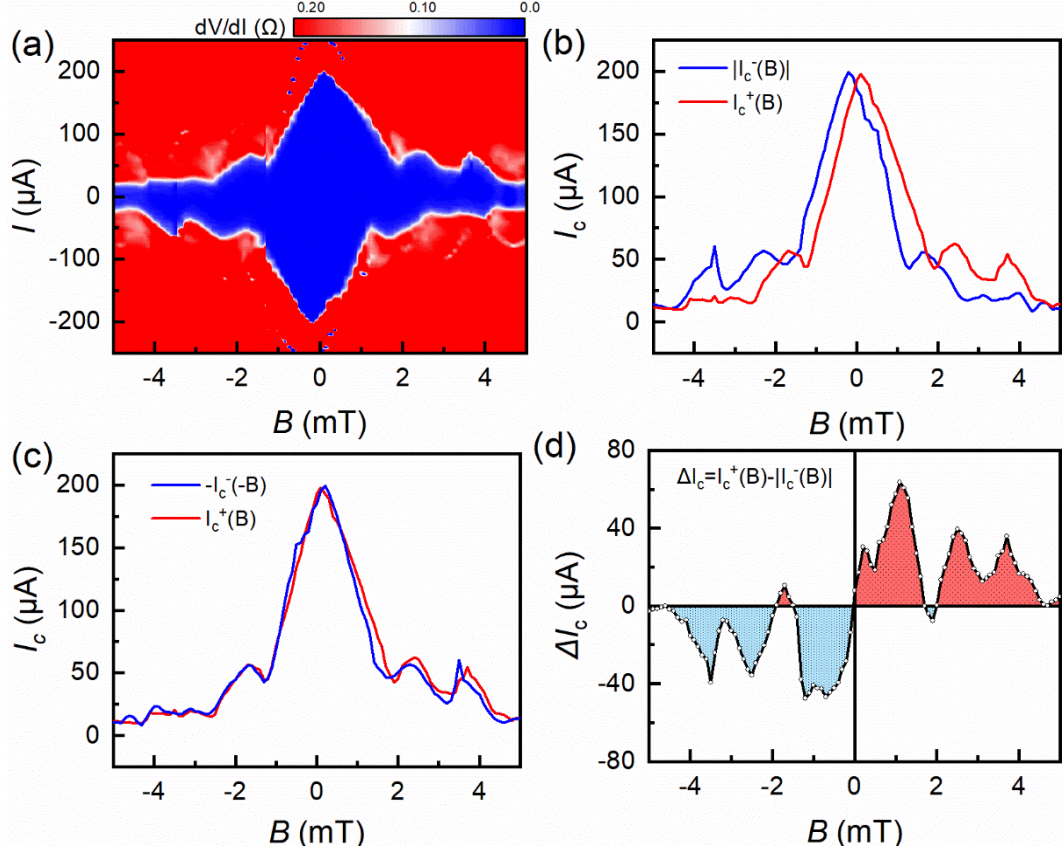

**Figure S3.** (a) Color map of the differential resistance versus the current bias and magnetic field at $T=0.5$ K for the Josephson junction with $L=500$ nm, $W=2$ μm, and $t=43$ nm. (b) Field dependence of the magnitudes of critical supercurrents in positive ($I_c^+$) and negative ($|I_c^-|$) current directions. (c) Antisymmetry of the Josephson supercurrent, *i.e.*, $|I_c^\pm(B)| = |I_c^\mp(-B)|$. (d) Field dependence of the asymmetry in critical supercurrents defined by $\Delta I_c = I_c^+(B) - |I_c^-(B)|$. The field regions with positive and negative signs of $\Delta I_c$ are shaded with the light red and blue colors, respectively.

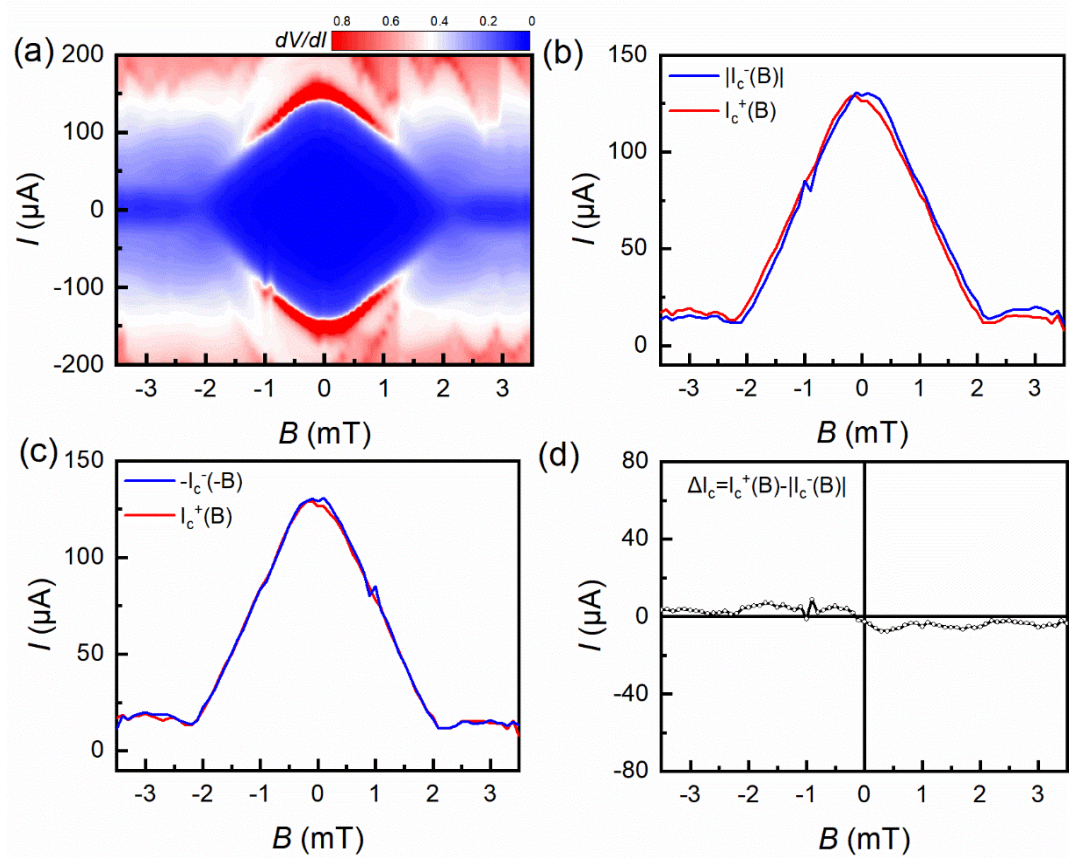

**Figure S4.** (a) Color map of the differential resistance versus the current bias and magnetic field at $T=0.1$ K for the Josephson junction with $L=500$ nm, $W=2$ μm, and $t=66$ nm. (b) Field dependence of the magnitudes of critical supercurrents in positive ($I_c^+$) and negative ($|I_c^-|$) current directions. (c) Antisymmetry of the Josephson supercurrent, *i.e.*, $|I_c^\pm(B)| = |I_c^\mp(-B)|$. (d) Field dependence of the asymmetry in critical supercurrents defined by $\Delta I_c = I_c^+(B) - |I_c^-(B)|$. Compared with Josephson junctions based on thinner MoTe$_2$ flakes, it exhibits almost negligible asymmetry.

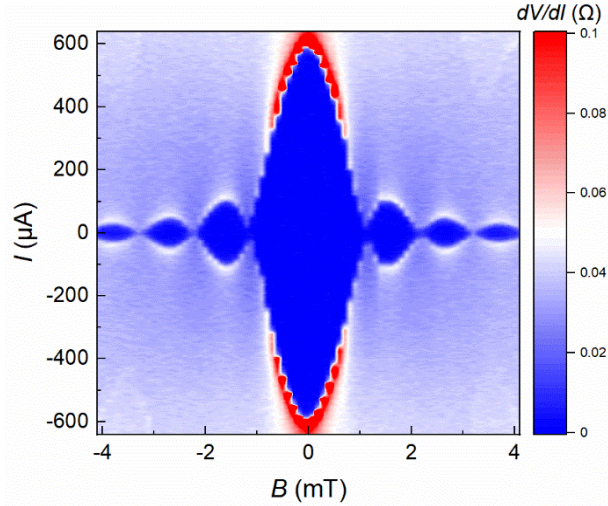

**Figure S5.** Color map of the differential resistance versus the current bias and magnetic field at $T=1.5$ K for a Nb/PdTe$_2$/Nb Josephson junction with $L=233$ nm, $W=2$ μm, and $t=265$ nm, exhibiting standard symmetric Fraunhofer pattern.

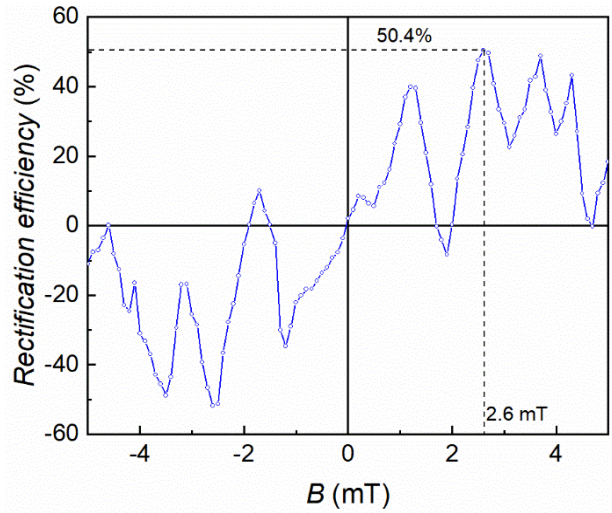

**Figure S6.** Field dependent rectification efficiency of MoTe$_2$-based superconducting diodes. The efficiency is defined by $\frac{I_c^+ - |I_c^-|}{I_c^+ + |I_c^-|}$. We calculate the efficiency based on the experimental results shown in Figure S3 (b) & (d). The maximum value of efficiency can reach up to 50.4% at 2.6 mT, as indicated in the figure.

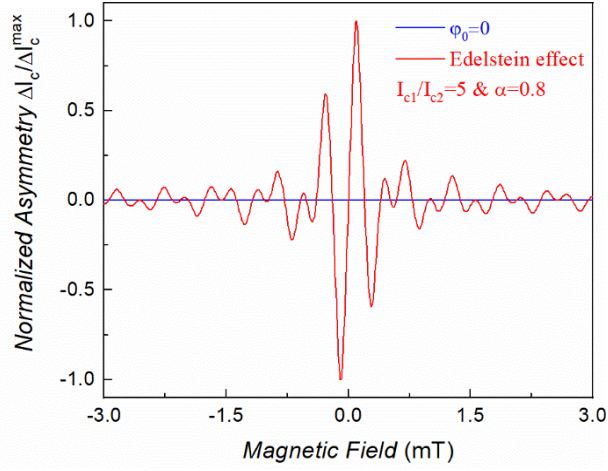

**Figure S7.** Critical supercurrent asymmetry $\Delta I_c/\Delta I_c^{max}$ numerically obtained by assuming the current phase relation $I_s = I_{c1}sin(\varphi + \varphi_0) + I_{c2}sin2\varphi$ with $\varphi_0 = \alpha B$. No AJE is observed for $\alpha=0$ or $\varphi_0=0$ (blue curve), but field tunable AJE appears once the Edelstein effect is taken into account (red curve).

## S1 Crystal growth and structure characterization

The MoTe$_2$ crystals were grown by the Te-flux method. Mo powder and Te blocks were first mixed and placed into a quartz ampoule. The ampoule was then heated up to 1273 K in 10 h and dwelled for 1 day. After that, the ampoule was cooled down to 1093 K at a rate of 1.5 K/h. Long flake-like crystals were obtained after centrifuging the excess Te flux. We have used the powder and single crystal X-ray diffraction techniques to confirm the $1T'$ phase of MoTe$_2$ at room temperature, as shown in Fig. S1. Note that the phase will transform to the $T_d$ phase at low temperatures as reported [1].

## S2 The Edelstein effect in noncentrosymmetric MoTe$_2$

The $T_d$ phase of bulk MoTe$_2$ possess a mirror plane perpendicular to the *a*-axis, a glide plane that is orthogonal to the *b*-axis and a two-fold screw axis that aligns with the *c*-axis. It is classified within the non-centrosymmetric space group Pmn2$_1$, which corresponds to the point group $C_{2v}$ [2]. According to the discussion of the magnetoelectric effect in gyrotropic superconductors [3], the magnetoelectric pseudotensor $T_{kj}$ for the $C_{2v}$ point group is $\begin{pmatrix} 0 & T_{xy} & 0 \\ T_{yx} & 0 & 0 \\ 0 & 0 & 0 \end{pmatrix}$. If we apply an in-plane supercurrent $J_j^S = \begin{pmatrix} J_x^S \\ J_y^S \\ 0 \end{pmatrix}$, the Edelstein-effect induced spin magnetization is given by $M_k = T_{kj}J_j^S = \begin{pmatrix} T_{xy}J_y^S \\ T_{yx}J_x^S \\ 0 \end{pmatrix}$, *i.e.*, $M_z = 0$. In perpendicular magnetic field, $\vec{M} \cdot \vec{B} = 0$. Therefore, we cannot observe the AJE in perpendicular fields for the point group $C_{2v}$. But a recent experimental study has revealed that due to the intrinsic confinement effect on the MoTe$_2$ stacking dependent free energy, disordered stacking occurs in thin MoTe$_2$ flakes, with only nanoscale $T_d$ domains [4]. For example, in a 37 nm thick MoTe$_2$ flake, $T_d$ domains with the average thickness of about 4 nm are observed using atomically resolved scanning transmission electron microscopy [4]. This suggests that the point group will be reduced to $C_{1v}$ for thin MoTe$_2$ flakes [3], which breaks the glide plane and screw axis symmetries. For the point group $C_{1v}$, the magnetoelectric pseudotensor $T_{kj}$ is given by $\begin{pmatrix} 0 & T_{xy} & 0 \\ T_{yx} & 0 & T_{yz} \\ 0 & T_{zy} & 0 \end{pmatrix}$. Therefore, an in-plane supercurrent would generate a spin magnetization $M_k = T_{kj}J_j^S = \begin{pmatrix} T_{xy}J_y^S \\ T_{yx}J_x^S \\ T_{zy}J_y^S \end{pmatrix}$, which has a nonzero $z$ component $M_z = T_{zy}J_y^S$. With the magnetic field applied perpendicularly to the plane, $\vec{M} \cdot \vec{B} \neq 0$. One would thus expect to observe the AJE or SDE in Josephson junctions based on thin MoTe$_2$ flakes.

The above theoretical expectation is in good1 agreement with our experimental results. As shown in Fig. 3 and supplementary Fig. S3 with the MoTe$_2$ layer thickness $t$=33 and 43 nm respectively, obvious AJEs are observed for the point group $C_{1v}$. But the AJE is greatly suppressed or almost disappears in supplementary Fig. S4 with $t$=66 nm. This is because thick MoTe$_2$ flakes tend to exhibit long range stacking order with increasing $t$ [4], eventually leading to a $C_{2v}$ point group symmetry as predicted for the bulk MoTe$_2$ [2].

## S3 Field-tunable asymmetric Josephson effect

The asymmetry $\Delta I_c$ in Fig. 4 exhibits oscillatory variation with increasing fields, with the sign of $\Delta I_c$ reversed and the magnitude changing non-monotonically with increasing fields. In order to fully understand this phenomenon, one needs to know the relation between the Edelstein-effect-induced anomalous phase $\varphi_0$ and the applied magnetic field $B$. Although $\varphi_0(B)$ has been well studied in both ballistic and diffusive JJs with Rashba-type spin-orbit coupling [5], no such theoretical calculation of $\varphi_0$ has been performed in JJs with the weak link material belonging to the $C_{1v}$ point group. But as a starting point, we have tried numerical simulations by simply assuming that the Edelstein-effect-induced anomalous phase $\varphi_0$ is proportional to the magnetic field $B$ and including the second harmonic in the current-phase relation (CPR) of JJs, i.e., the CPR is given by $I_s = I_{c1}sin(\varphi + \varphi_0) + I_{c2}sin2\varphi$ [6]. $\varphi$ is the phase difference between the two superconducting electrodes which can be tuned by the magnetic flux penetrating through the junction and $\varphi_0 = \alpha B$, accounting for the Edelstein effect. The preliminary results are shown in Fig. S7, Supporting Information. Without the Edelstein effect, i.e., $\varphi_0$=0, no asymmetry $\Delta I_c$ is observed, as indicated by the flat blue curve. After we take into account the Edelstein effect ($\alpha \neq 0$), the asymmetry appears. As can be seen from the red curve, the obtained $\Delta I_c(B)$ curve is antisymmetric and exhibits clearly oscillatory variations with the field, similar to the experimental results shown in Fig. 4. These preliminary results are quite encouraging and provide more support to the Edelstein-effect-induced AJE in MoTe$_2$ JJs.

# Supporting References